\documentclass[12pt,preprint]{aastex}

\slugcomment{ApJ in press}

\shorttitle{AGN Reddening and Continuum Shapes}
\shortauthors{Gaskell, Goosmann, Antonucci \& Whysong}

\begin{document}

\title{THE NUCLEAR REDDENING CURVE FOR ACTIVE GALACTIC NUCLEI AND
THE SHAPE OF THE INFRA-RED TO X-RAY SPECTRAL ENERGY DISTRIBUTION}

\author{C. MARTIN GASKELL and REN\'{E} W. GOOSMANN \altaffilmark{1}}
\affil{Department of Physics \& Astronomy, University of Nebraska,
Lincoln, NE 68588-0111} \email{mgaskell1@unl.edu}
\email{rene.goosmann@obspm.fr}

\and

\author{ROBERT R. J. ANTONUCCI and DAVID H. WHYSONG} \affil{Department of
Physics
\& Astronomy, University of California, Santa Barbara, CA 93106}
\email{ski@spot.physics.uscb.edu} \email{dwhysong@spot.physics.uscb.edu}

\altaffiltext{1}{Present address: Observatoire de Paris, Section
de Meudon, LUTH, Place Jules Janssen, F-92195 Meudon Cedex,
France}

\begin{abstract}

We present extinction curves derived from the broad emission lines
and continua of samples of 72 radio-loud and 1018 radio-quiet
AGNs. The curves are significantly flatter in the UV than are
curves for the local ISM. The reddening curves for the radio-quiet
LBQS quasars are slightly steeper than those of the radio-loud
quasars in the UV, probably because of additional reddening by
dust further out in the host galaxies of the former. The UV
extinction curves for the radio-loud AGNs are very flat. This is
explicable with slight modifications to standard MRN dust models:
there is a relative lack of small grains in the nuclear dust. Our
continuum and broad-emission line reddening curves agree in both
shape and amplitude, confirming that the continuum shape is indeed
profoundly affected by reddening for all but the bluest AGNs. With
correction by our generic extinction curve, all of the radio-loud
AGNs have continuous optical-UV spectra consistent with a single
shape. We show that radio-quiet AGNs have very similar intrinsic
UV to optical shape over orders of magnitude in luminosity. We
also argue that radio-loud and radio-quiet AGNs probably share the
same underlying continuum shape and that most of the systematic
differences between their observed continuum shapes are due to
higher nuclear reddening in radio-selected AGNs, and additional
reddening from dust further out in the host galaxies in
radio-quiet AGNs. Our conclusions have important implications for
the modelling of quasar continua and the analysis of quasar
demographics.

\end{abstract}


\section{INTRODUCTION}

Poor characterization of the extinction in Active Galactic Nuclei
(AGNs) has been one of the biggest obstacles to understanding
their nature.  Without knowledge of the appropriate form and
amplitude of the wavelength dependence of the extinction it is
impossible to determine the true spectral energy distribution
(SED). It is also difficult to interpret emission-line spectra to
diagnose conditions nears the black hole.  There has been a
long-standing controversy over the amount of reddening of AGNs.
McKee \& Petrosian (1974) argued against significant amounts of
reddening in quasars on two grounds.  The first was the apparent
lack of the strong $\lambda2175$ extinction hump (e.g., Pitman,
Clayton, \& Gordon 2000), and the second was the lack of curvature
of the UV continuum. Cheng, Gaskell, \& Koratkar (1991) found that
the ultraviolet slopes for a wide variety of AGNs at high and low
redshift, high and low luminosity, and differing radio types were
essentially the same to within their measuring errors.  They
limited line-of-sight reddening to E(\bv) $ \sim 0.1$ mag
(assuming a Galactic reddening curve). In the optical, on the
other hand, there is clear evidence at least for significant
reddening in the host galaxy plane. De Zotti \& Gaskell (1985)
showed that both the lines and continua of radio-quiet AGNs
(Seyfert galaxies) with axial ratio $a/b$ are typically reddened
by E(\bv) $\sim 0.2(a/b)$ mag.  Also, unified models of both
radio-loud and radio-quiet AGNs require significant dust
extinction in a nuclear torus (tipped with respect to the spiral
host; see Antonucci 1993) and this must produce detectable effects
on spectra of both radio-loud and radio-quiet AGNs.

Reddening curves have been determined for the radio-quiet AGNs
NGC~3227 (Crenshaw et al. 2001), and Ark 564 (Crenshaw et al.
2002) by assuming that they have the same underlying continuum as
other, bluer AGNs. These reddening curves show little or no
$\lambda2175$ bump and rise {\em more} steeply in the far UV than
the standard Galactic reddening curve (i.e., they resemble the
reddening curve of the SMC). Crenshaw et al. (2001, 2002) argue
that the dust in NGC~3227 and Ark 564 is outside the narrow-line
regions and hence far removed from the nuclei. Similarly, Richards
et al. (2003) show that SMC-like reddening affects the spectra of
a few percent of radio-quiet quasars.  In this paper we determine
mean reddening curves for large samples of radio-loud and
radio-quiet quasars, and argue that {\it most} quasars are also
affected by substantial extinction which has a quite flat
reddening curve in the UV. This curve is qualitatively different
from any previously proposed reddening law. We explore the
significance of this result for our understanding of AGNs.

\section{THE REDDENING CURVE}

\subsection{The Sample}

Baker \& Hunstead (1995) present composite Anglo-Australian
Telescope optical and (rest-frame) UV spectra of 72 FR II radio
quasars and broad line radio galaxies from the Molonglo Quasar
Sample (MQS). The data have the advantages of being homogenous and
being selected almost exclusively by radio-lobe flux, a nearly
isotropic parameter.  They also have broad optical/UV wavelength
coverage, typically over a rest-frame wavelength range of $\sim
2000$ \AA\ to $\sim 5000$ \AA, depending on redshift. Radio
core-to-lobe flux ratios, $\Re$, were measured from 5 GHz VLA
maps. Baker \& Hunstead excluded a few AGNs from the sample
because of abnormally strong absorption features, a lack of
$\Re$-values, or a BL Lac type featureless spectrum. They grouped
the AGNs into four subsets chosen according to $\Re$-values and
radio structure, creating composite UV-optical spectra of 13-18
objects each.  The four subsets are $\Re \geqslant 1$, $0.1
\leqslant \Re < 1$, $\Re < 0.1$ and Compact Steep Spectrum (CSS)
radio sources.  Baker \& Hunstead (1996) also give mean line
strengths for the four subsets.  In addition to the MQS AGNs we
also considered the similar quality predominantly radio-quiet
Large Bright Quasar Survey (LBQS) composite spectrum of Francis et
al (1992), based on the spectra of 1018 quasars.

\subsection{Method and Justification of Method}

Our continuum extinction curves are derived by division of pairs
of composite continuum spectra, on the assumption that the
intrinsic continua are the same. While this is routinely done for
stars of a given spectral type, there is in principle a huge risk
in applying this method to AGN.  Stars with matched
classifications almost certainly have intrinsically similar
spectra, but there is no guarantee that one AGN composite is
intrinsically similar to another.  In fact for the AGN case,
variability appears to affect the spectral shape so the intrinsic
matches cannot be perfect for that reason alone.  However, the
averaging within the composites will greatly reduce the last
problem. Note that our sub-samples are matched in radio-lobe
power.

We also determined the reddening curve for the broad-line region
(BLR) lines in the samples.  In doing this we are, again, assuming
that there are no intrinsic object-to-object BLR differences. This
is certainly not true for individual objects but again we hope
that the large number of AGNs in each subset minimizes the
uncertainties introduced.

There are two very good pieces of evidence we will present that
reddening (rather than intrinsic spectral differences) does in
fact drive most of the observed spectral differences.  The first
is that a {\em single} reddening curve shape reconciles the shapes
all four radio composites.

More importantly, we shall see that {\em the reddening curves
derived from the continua and separately from the broad emission
lines have nearly the same shape and amplitude.}  Changing
physical conditions in the BLR  cannot mimic reddening when many
lines are considered together \footnote{However, a single line
pair alone often gives ambiguous results -- see Grandi (1982).}.

As explained at the end of this paper, the validity of our
conclusion will soon be tested in a simple robust
manner: it predicts powerful infrared emission for the
lobe-dominant radio-loud quasars, far exceeding that of the
blue-selected PG radio-quiet quasars.

\subsection{Results}

\subsubsection{Choice of Comparison Groups}

 There are four radio quasar samples considered in the paper:
core-dominant classical doubles, intermediate classical doubles,
lobe-dominant classical doubles, and compact-steep-spectrum
sources. Roughly speaking, the current wisdom is that the first
three classes differ from one another only in orientation, while
the last class comprises sources which are intrinsically
different, in particular being much smaller physically.

We concentrate on two comparisons among the groups.  First, we
compare the core-dominant versus the lobe-dominant sample.  There
is empirical and theoretical evidence that the latter are more
reddened, so such a comparison should reveal the shape of that
reddening.

The compact-steep-spectrum quasars are thought to range widely in
orientation. They fill the parameter (viewing angle) space of all
three groups of classical double radio quasars considered
above\footnote{Note that the {\it edge-on} quasars are thought to
be optically-obscured and to manifest themselves as
radio-galaxies.} Thus, if we are to compare them to a subgroup of
the classical doubles, it would be best to select the subgroup at
intermediate orientation as we have done. One might compare them
to a composite of all the classical doubles, but restricting the
comparison group to those at intermediate orientation has another
very important advantage: our two comparisons and thus our two
reddening curves for the radio loud population are strictly
independent of each other, with no objects in common. Thus their
mutual agreement cannot be an artifact of interdependent samples.

\subsubsection{Reddening Curve Comparisons}

We derived relative reddening curves for two independent pairs of
sub-samples of the four composite spectra given by Baker \&
Hunstead.  We compare the $\Re \geqslant 1$ (``face-on'') sample
with the $\Re < 0.1$ (``edge-on'') sample, and the $0.1 \leqslant
\Re < 1$ (``intermediate-orientation'') sample with the CSS
sample. We avoided prominent spectral features (like major broad
or narrow emission lines and absorption troughs) or, in some
cases, we made interpolations over residual absorption features
and noisy regions of the spectra.  We also determined the
extinction curves for the BLR using the mean line strengths given
in Table 1 of Baker \& Hunstead (1996a,b). Because of the
difficulty of measuring BLR strengths due to the wings of the
lines the uncertainties are larger for the BLR lines than for the
continua. In Figure 1 we show each of the $E(\lambda-V)$ curves
normalized to $E(\bv) = 1$.  Of course the uncertainty in the
normalization in the optical introduces some uncertainty in the
level of the UV curves.

The two continuum extinction curves, coming from the two totally
independent data subsets, agree well, especially since we have
only matched the curves in the optical region, rather than using a
globally-optimal scaling to get the best overall match.  It is
important to note that the BLR extinction curve is also consistent
with the two continuum curves\footnote{It should be noted that our
scenario therefore allows us to explain much of the high observed
ratios of Balmer line strengths to Ly $\alpha$ by extinction.
These ratios are still dependent on physical conditions in the BLR
however, as is shown by variations of the Balmer to Ly $\alpha$
line ratios across line profiles (see Snedden \& Gaskell 2004).}.

All three reddening curves are extremely flat in the UV. This is
very different behavior from the standard Galactic reddening curve
(with a ratio of total to selective extinction, $R_V = 3.1$) shown
for comparison in Figure 1. It is also very different from the
Small Magellanic Cloud (SMC) reddening curve that has also been
considered as a possible form of the wavelength dependence of the
extinction in AGNs (Crenshaw et al. 2001, 2002). Reddening by a
galactic curve or an SMC curve turns a power-law into a convex
spectrum (see McKee \& Petrosian 1974) while our nuclear reddening
curve produces a concave spectrum. As noted in the Introduction,
Richards et al. (2003) find that a few percent of the SDSS (mostly
radio-quiet) AGN are significantly affected by SMC-like reddening
and our results are consistent with that. Richards et al. state
that attempting to explain their large color-segregated composites
generally with reddening ``results in good matches at both
1700\AA\ and 4040\AA, but over predict(s) the flux between these
two wavelengths and under predict(s) the flux shortward of C IV.''
This is exactly what is predicted by our reddening curve in Fig.
1, and qualitatively confirms that our picture appears to be very
broadly applicable.

A flat extinction curve implies large grain sizes relative to the
wavelengths involved. We note in fact that Galactic reddening
curves from regions having a large value of the total extinction
to selective extinction ratio, $R_V \sim 5$, such as the line of
sight to the star formation region Herschel 36 (Fitzpatrick \&
Massa 1988), do give flat UV extinction curves. The major
difference between the Herschel 36 reddening curve and our
radio-loud quasar curves is the absence of the $\lambda2175$
absorption feature in our curves. For comparison, we show a
parameterized extinction curve with $R_V = 5.30$ (a good fit to
the Herschel 36 line of sight), and a ``normal'' extinction curve.
Both curves are taken from Cardelli, Clayton \& Mathis (1989,
hereinafter CCM89).

Next, we consider the predominantly radio-quiet (and
blue-selected) LBQS survey objects.  Figure 2 shows the relative
reddening curve between the composite spectrum from Francis et al.
(1992) and the relatively unreddened $\Re \geqslant 1$ composite
of Baker \& Hunstead (1995).  This shows the excess reddening
relative to this radio-loud sample.  Since the LBQS composite
spectrum is rather noisy in the V-band, we did the normalization
to $E(V-V) = 0$ and $E(\bv) = 1$ by doing a fit to the four data
points with lowest frequencies. The resulting reddening curve is
significantly steeper than the radio-loud reddening curves in
Figure 1, but not quite as steep as the Galactic reddening curve,
or the NGC~3227 and Ark~564 reddening curves of Crenshaw et al
(2001, 2002). Again, there is the lack of the $\lambda2175$ bump.
In fact, there is an apparent dip in the reddening curve around
the $\lambda2175$ bump. This would be hard to understand
physically unless there were a scattering component to the
{$\lambda2175$} extinction feature. We suspect instead that this
is due to a break down of our assumption that the intrinsic
continua are the same for the samples. This could be a consequence
of a difference in the UV Fe~II emission. Part of the well-known
Boroson \& Green (1992) ``eigenvector-1'' correlations (which
include differences between radio-loud and radio-quiet AGNs) is
variation in the Fe~II emission, and the Fe~II emission is
important in the so-called ``small blue bump'' (SBB) continuum
around $\lambda^{-1} \sim 3.5$.  De Zotti \& Gaskell (1985)
discuss the effects of Fe~II on measurement of a $\lambda2175$
absorption feature.  Note that in deriving the reddening curves
there is no reason to exclude SBB emission, which arises in the
BLR; the dust affects it the same way as the BLR and continuum
source.  While $\lambda2175$ has almost always been reported to be
weak in AGNs, it should be noted that De Zotti \& Gaskell (1985)
did find a correlation between their estimated reddening and
published strengths of the $\lambda2175$ feature (see their Table
3 and their Fig. 12) for dust in the host galaxy plane of
Seyferts.  There are also occasional reports (e.g., Grossan et al.
1996) of a significant $\lambda2175$ feature in quasars.

\subsection{The Ratio of Total to Selective Extinction ($R_V$)}

We estimated $R_V$ for our extinction curves by two methods. First
we followed the standard approach of extrapolating $E_\lambda$ to
$\lambda^{-1} = 0$. Actual measurements of $E_\lambda$ in the IR
are not available and would give inaccurate results anyhow because
of the onset of IR emission from dust.  Therefore, following the
procedure of CCM89, we extrapolated to $\lambda^{-1} = 0$ by
assuming that the IR reddening curve has the same shape as the
Galactic curve of Rieke \& Lebofsky (1985).\footnote{The IR
reddening is relatively low, and known reddening curves differ
only slightly in that region, so this is probably safe} To match
Rieke \& Lebofsky (1985) we converted our curves to
$A_\lambda/A_V$ using the relation:

\begin{math}
  A_\lambda/A_V  = \frac{A_\lambda-A_V}{A_V} + 1 = \frac{A_\lambda
  -A_V}{R_V \cdot E(B-V)} + 1 = \frac{1}{R_V} \cdot
\frac{E(\lambda-V)}{E(B-V)} + 1.
\end{math}

We show one of our radio-loud $A_\lambda/A_V$ AGN extinction
curves together with the infrared data of Rieke \& Lebofsky (1985)
in Fig. 3. We allowed the Galactic curve to vary by a scale factor
$C$ (see CCM89). By assuming a similar shape for the AGN and
Galactic sample, we obtain $R_V$.

CCM89 show that the main variations in Galactic optical-to-UV
extinction laws are a function of $R_V$ alone and this gives
another way to find an estimate of $R_V$. In order to use the
CCM89 relationships for our extinction curves, we had to modify
their relationships for the optical/near-UV branch to account for
the lack of the $\lambda2175$ feature. For $3.3\mu m^{-1} \leq x
\leq 8 \mu m^{-1}$ with $x=\lambda^{-1}$ we hence defined the
function $\bar b(x)$ as

\begin{math}
  \bar b(x) = -3.09+1.825x+1.5/[(x-4.62)+0.85]+F_b(x)
\end{math}

instead of function 4b in CCM89. We adjusted $R_V$ in the modified
CCM89 approximations until they fit our extinction curves.  From
this we get two estimates of $R_V$, one from the fit to the
optical/near-UV range and the other from the fit the far-UV
branch. We summarize our resulting $R_V$ values in Table 1.

Within our estimated errors, all three wavelength ranges (IR,
UV/optical, and far-UV) lead to similar estimates of $R_V$, and
all $R_V$ values for our radio-loud curves are significantly
greater than the canonical Galactic value of 3.1.

For NGC 3227 Crenshaw et al. (2001) obtained $R_V = 3.2$ and for
Akn 564 Crenshaw et al. (2002) obtained $R_V  = 3.1$. Both of
these are very close to the Galactic value, but these Seyfert
curves likely refer to a non-nuclear environment.

Some additional support for the flat shape of our reddening curves
in the UV comes from the lack of luminosity dependence of UV
slopes found by Cheng et al. (1991), and the sharp upper limit to
the $\lambda4220/\lambda1460$ ratio found by Malkan (1984). We
discuss this in section 4.

\section{THE CONTINUUM SHAPE AND DEGREE OF REDDENING IN AGNS}

In deriving the extinction curves for the radio quasars in the
previous section, we started by assuming that the
optical/UV-continuum shapes were the same \emph{for each pair} of
composites.  This, by construction, leads to inferred intrinsic
spectra which match each other for each pair.  But what is {\em
not} true by construction, and what argues for a rather generic
reddening law, is that the two continuum extinction curves,
derived from independent data, agree with each other in shape, as
can be seen in Figure 1. More remarkably, as can also be seen in
Figure 1, the continuum reddening curves for each pair agree in
both shape and amplitude with curves derived from the {\em
emission lines}. Again, this is not true by construction, but
argues for a correct and generic extinction law. To summarize:
both independent continuum pairs lead to the same extinction
curve, and the corresponding emission lines produce an extinction
curve which also agrees in shape and amplitude.

There is already evidence for a universal spectral energy
distribution in radio-quiet AGN.  In Fig. 4 we show the
$\lambda1600$/$\lambda 4220$ spectral index, $\alpha_{UVO}$ (for
$F_{\nu} \propto \nu^{-\alpha}$), for radio-quiet AGNs in three
optical luminosity bins in $L_{4200}$ (in erg sec$^{-1}$
Hz$^{-1}$). We have calculated these from data in Malkan (1984)
and references therein.  We only show spectral indexes for objects
with simultaneous or nearly simultaneous UV and optical
observations. The observations were already corrected for Galactic
reddening using $E(\bv)$ values from Burstein \& Heiles (1982) and
assuming a standard Galactic reddening curve.

For the highest luminosity sample ($ \vert $log$L \vert > 30.5$),
the mean $\alpha_{UVO}$ is 0.54, and the standard deviation is
only 0.064. If this small dispersion is due to reddening, it
corresponds to E(B-V) $\sim 0.06$, assuming a reddening curve like
our Fig. 1, or to E(B-V) $< 0.02$, if we assume a standard
Galactic reddening curve. Thus, as was argued by Malkan (1984),
the intrinsic extinction in these high-luminosity AGNs is probably
very low.

For the radio-quiet AGNs with $29 < \vert $log$L \vert < 30.5$,
the important thing to note is that the {\it cutoff at}
$\alpha_{UVO} \sim 0.45$ {\it is the same as for the highest
luminosity AGNs}\footnote{Of course, our slopes cannot be compared
directly with published values defined differently.  For example,
Richards et al. (2003) quote ``photometric'' spectral indices,
measured using fixed observed-frame optical bands, with only
statistical corrections for emission and absorption features. Many
widely varying spectral index prescriptions are used in the
literature.}. Both reddening and contamination by starlight from
the host galaxy will increase $\alpha_{UVO}$.  For $\vert $log$L
\vert < 29$ we see again that \emph{there is the same cutoff at
$\alpha_{UVO} \sim 0.45$}. As noted by Malkan (1984), the
constancy of the cutoff slope in the face of a difference in the
mean slope can be totally explained by increasing the host galaxy
contamination and reddening.  However, when we compare reddenings
from the optical and UV (see below), we will see that reddening,
not host galaxy contamination is the dominant factor. The cutoff
at $\alpha_{UVO} \sim 0.45$ is presumably the unreddened value.
{\em It is remarkable that this is the same over more than 4
orders of magnitude in luminosity.} It would be hard to explain
the increasing spread in spectral index with decreasing luminosity
as an intrinsic luminosity effect given that the {\em cutoff} is
independent of luminosity.

Ward et al. (1987) argue that all AGN continua are the same and
that differences are due to reddening and host galaxy starlight
contamination.  For low-redshift, radio-loud quasars, Netzer et
al. (1995) showed that there is a correlation between the observed
$Ly\alpha/H\beta$ line ratios and the $\lambda1216 / \lambda4861$
continuum flux ratios. This is evidence for radio-loud quasars
having similar unreddened continuum shapes and similar
hydrogen-line ratios. The bluest $\lambda1216/\lambda4861$ flux
ratio on the reddening vector in their Fig. 6 is 8.7. This implies
$\alpha_{UVO} \sim 0.44$ and supports the idea of a universal
$\alpha_{UVO} \sim 0.45$.  However, Bechtold et al. (1996) found a
similar correlation (for the same wavelength range) for
high-redshift predominantly radio-quiet AGNs but the blue end of
their reddening vector gives $\alpha_{UVO} \sim 0.2$.

Assuming that the bluest AGNs have $\alpha_{UVO} = 0.45$, and
adopting our radio-loud-AGN reddening curve, the mean reddenings
(E(B-V)) of the four Molongo samples, going from ``face-on'' to
CSS are, 0.29, 0.34, 0.71, and 0.98 mag. (with formal relative
uncertainties of about $\pm 0.06$).  If the ``face-on'' sample is
assumed to be unreddened then the mean reddenings are 0.00, 0.05,
0.41, and 0.69 magnitudes respectively, but this requires the
unreddened $\alpha_{UVO} = 0.70$.  We consider this highly
unlikely since we get flatter (bluer) {\em observed} spectral
indices from the Netzer et al. (1995) measurements. Since the
reddening curve for the predominantly radio-quiet LBQS sample
appears to be different, we derived the reddening from the long
wavelength region (i.e., the optical) of the spectrum only.  This
gives a mean E(B-V) $\sim 0.30$ mag (i.e., similar to the face-on
Molongo radio-loud AGNs) if we assume that the unreddened
$\alpha_{UVO} = 0.45$.  Ward \& Morris (1984) got E(B-V) $\sim
0.3$ for NGC~3783; Crenshaw et al. (2001) got a total E(B-V) of
0.20 for NGC~3227; and Crenshaw et al. (2002) got a total E(B-V)
of 0.17 for Ark 564.  All three of these are in the range
considered here.

Needless to say, the lower apparent reddenings of radio-quiet, and
hence, usually optically-selected, AGNs must be strongly
influenced by selection effects.  It would be interesting to study
the reddenings of X-ray selected samples of radio-quiet AGNs.

It should be noted that after de-reddening with our reddening
curve, there is no difference between the spectral energy
distribution of CSS sources and other AGNs.  The apparent weakness
of the SBB in CSS sources is simply a consequence of the high
degree of reddening and the shape of the reddening curve which has
its greatest curvature at the position of the SBB. Our reddening
curve might similarly explain the unusual spectral shapes of some
broad-absorption line quasars without the need to invoke partial
covering (see Hall et al. 2002).

\section{LUMINOSITY DEPENDENCE OF REDDENING}

Assuming, as we have argued in the previous section, that the
intrinsic continuum really is independent of luminosity, we can
use our nuclear reddening curve to investigate the luminosity
dependence of the reddening.  In Fig. 5 we show the luminosity
dependence of the reddenings, calculated from the UV to optical
slope (as discussed in the previous section) and from the optical
$\lambda$4200 to $\lambda$7000 slopes given by Mushotzky \& Wandel
(1989).  It should be noted that the Malkan (1984) and Mushotzky
\& Wandel (1989) samples are two different heterogenous samples.
We have assumed that the spectra are unreddened for the highest
luminosities.  The error bars are the errors in the means
($\sigma$/$\sqrt(n)$.)

The first thing to notice from Fig. 5 is that the reddenings,
derived by the two methods, and from the two different samples,
are in rough agreement.  The differences in the reddenings deduced
by the two methods for the two lowest luminosity bins are
comparable to the scatter in the reddening curve in Fig. 1.
Although the differences exceed the formal $\sigma$/$\sqrt(n)$
error bars, we consider the two derivations to be in satisfactory
agreement.  There would not be this level of agreement had we used
a reddening curve which is more selective in the UV.  For example,
had we used the standard Galactic reddening curve, the reddenings
derived from the UV-optical slope (the solid squares) would then
be more than a factor of two lower. This suggests that our
proposed nuclear reddening curve is more realistic.

It is important to recognize that the agreement in reddenings
deduced from the two spectral regions would also not occur if the
steepening of spectral indices at lower luminosities were due to
increasing host galaxy contamination. To illustrate this, in Table
2 we show the reddenings, E(B-V), that would be inferred using our
reddening curve if host galaxy contamination were interpreted as
reddening. We approximated an unreddened AGN continuum as an
$\alpha$ = 0.45 power-law and added in host galaxy contamination
assuming the composite spiral galaxy SED in Table 6 of Schmitt et
al. (1997). Using their elliptical SED makes no significant. We
then interpreted the changes in flux ratios as a reddening. The
first column in Table 2 shows the ratio of AGN to galaxy flux at
7000 \AA. The last two columns give the corresponding reddenings
inferred using our reddening curve.

As can be seen from Table 2, if host galaxy contamination were to
be mistaken for reddening, the reddenings inferred from the
UV-optical would always be greater than those inferred from the
optical alone. Host galaxy contamination would produce an offset
between the optical-reddening and optical-only reddening in the
{\em opposite} sense to the slight difference shown in Fig. 5. Two
factors are responsible for host galaxy contamination leading to
greater apparent reddenings from the optical than from the
optical-UV. First, the SED of a host galaxy is steeper from the
optical to the UV than it is in the optical alone.  Second,
similar changes in UV-optical color and optical color correspond
to a larger change in the E(B-V) deduced from the UV-optical color
because our reddening curve is flatter from the optical to the UV
than it is in the optical region alone.

The differences in Fig. 5 between the reddenings deduced from
these two spectral regions could be due to departures from our
reddening curve (e.g., because of reddening by additional
extra-nuclear dust), but they could equally well be due to
differences in the samples.  The Malkan (1984) sample is
predominantly optically selected, and hence biased towards lower
reddening objects; the larger Mushotzky \& Wandel (1989) includes
a large fraction of radio-loud (and thus radio-selected) AGNs.

The second thing to notice from Fig. 5 is that the reddenings
derived by both methods increase monotonically with decreasing
luminosity. Mushotzky \& Wandel (1989) attributed the steepening
of the optical slope to intrinsic differences in the spectral
shapes, perhaps due to differing black hole masses and accretion
rates (Wandel \& Petrosian 1988). However, using our reddening
curve, we can accurately predict the luminosity dependence of the
optical slope in Mushotzky \& Wandel (1989) entirely from the
UV-optical reddening.  We are not ruling out the possibility that
there are luminosity dependencies in the intrinsic (de-reddened)
continuum shape.  Before these can be found, we must account for
the dominating effect of reddening.  Another prediction is that if
the dust has a reddening curve similar to that in Fig. 1, then the
lowest luminosity objects should show {\it concave} SEDs going
from the near IR to the far UV.

Variations in the degree of reddening from object to object
naturally explain the spread in continuum shapes at a given
luminosity.  In Fig. 4 it can be seen that as the mean spectral
index increases, so does its spread.  If the continua are
unreddened and the differences in continuum shape are intrinsic,
perhaps due to differences in black hole mass and accretion rate,
then the wide range of spectral indices at lower luminosity is
hard to understand.

\section{THE RELATIONSHIP BETWEEN THE X-RAY AND OPTICAL EMISSION}

\subsection{Luminosity dependence of $L_X/L_{opt}$}

It is long been known that the observed ratio of X-ray to optical
luminosity is a decreasing function of optical (or UV) luminosity
(Reichert et al. 1982, Zamorani et al. 1982, Avni \& Tananbaum
1982, Kriss \& Canizares 1985) with $L_X \propto
L_{opt}^{0.70-0.85}$. This has generally been attributed to
luminosity-dependent physical conditions in the AGNs themselves.
In contrast to the relationship between $L_X$ and $L_{opt}$,
Malkan (1984) showed that $L_X \propto L_{IR}$. Mushotzky \&
Wandel (1989) reconciled these results by showing that the
dependence of $L_X$ on the luminosity at longer wavelengths
depends on where the longer wavelength luminosity is measured.  We
believe instead that the $L_X/L_{opt}$ ratio appears to vary
simply because the greater extinction in lower luminosity objects
causes the optical luminosity to appear to be lower than it really
is, while X-ray luminosities at $\gtrsim 2$keV are unaffected.  We
quantitatively tested this by using $L_X \propto L_{opt}^{a}$, to
predict the observed optical luminosity, then assuming that the
intrinsic optical luminosity was proportional to $L_X$, and using
this to predict the reddening using our reddening curve.  The
luminosity dependence of the reddening seen in Fig. 5 is well
matched with $a \sim 0.8$. Going in the other direction, and
assuming the luminosity dependence in Fig. 5, we predict that
there should be some curvature in the $L_X$ vs. $L_{opt}$ plot.
This should be detectable with well-defined samples.

\subsection{Dependence of $L_X/L_{opt}$ on Radio-loudness}

Zamorani et al (1982) reported that radio-loud AGNs are about
three times as luminous in X-rays as radio-quiet AGNs of the same
observed optical luminosity.  According to the arguments of the
present paper, much of this difference must result from a
denominator in $L_X/L_{opt}$ reduced by dust absorption.  A factor
of three corresponds to a difference in E(B-V) of $\sim 0.25$ --
0.35 (depending on $R_V$).  This is similar to the
radio-loud/radio-quiet differences we find in optical absorption
using our reddening curve (see above).

This is not the whole story however, because Zamorani et al
(1982), Kembhavi (1994), Shastri (1997), and Brinkmann et al 1997,
among others, have argued that there is at least some beamed
contribution to the $L_X/L_{opt}$ difference, because the
numerator is enhanced by a beamed component in the radio-loud
objects with strong cores. The clinching argument here is that
core-dominant objects often have harder X-ray spectra (Brinkmann
et al. 1997), heralding the arrival of a new spectral component,
but this observational effect is modest in size and ubiquity.  We
note that Sambruna, Eracleous, \& Mushotsky (1999) find that X-ray
slopes are similar for radio-loud and radio-quiet AGNs when the
samples are matched in luminosity.

A quantitative assessment of the relative importance of X-ray
beaming and optical absorption to the enhanced $L_X/L_{opt}$ in
radio-loud AGNs is far beyond the scope of this paper.  We simply
state here that, if we are correct, a significant part of the
higher average $L_X/L_{opt}$ in radio-loud AGNs must be due to a
denominator diminished by optical absorption in many cases, and
this must be accounted for as well as the beaming.

\section{DISCUSSION}

\subsection{Evidence for Larger Grain Sizes}

The simplest explanation of the flat UV extinction curves we find
is that small grain sizes are depleted relative to the grain-size
distribution in our galaxy. There is other evidence for depletion
of small grains. It has long been known that the optical
extinction in AGNs is considerably less than is predicted by the
X-ray column densities (Maccacaro, Perola \& Elvis 1982, Reichert
et al. 1985). By comparing the reddening of optical and infrared
broad lines and the X-ray absorbing column density Maiolino et al.
(2001b) find that the $E(\bv)/N_H$ ratio is nearly always lower
than the Galactic values by a factor ranging from $\sim 3$ up to
$\sim 100$. They argue that this cannot be due to conversion of
refractory elements to the gas phase but is instead caused by
larger grains.  Sambruna et al. (2002) find that approximately
50\% of BLRGs have columns of cold gas comparable to the columns
detected in NLRGs.  This suggests that the small grains are
depleted in gas swept off the torus.

The silicate feature at $9.7 \mu m$ observed in the mid-IR spectra
of many Galactic sources is absent in the average ISO spectrum of
a sample of Seyfert 2 galaxies obtained by Clavel et al. (2000).
Maiolino, Marconi, \& Oliva (2001a) conclude that the most likely
explanation of this is that dust in the circumnuclear region of
AGNs is predominantly composed of large grains that do not
contribute to the feature at 9.7 $\mu$m.  Imanishi (2001) has
argued from the probable shape of the IR L and M' band extinction
curve that it is due to large grains.

There are other good arguments that small grains are severely
depleted in gas swept off the torus.  On the observational side,
we are confident that free electrons sometimes dominate the
scattering above the torus (Miller, Goodrich, \& Mathews 1991;
Ogle et al 2003). Small grains have orders of magnitude greater
scattering efficiency per gram of ionized Galactic interstellar
medium compared with electrons, so they must be very thoroughly
eliminated for electrons to dominate. In section 6.3 we discuss
ways in which small grains can be destroyed close to an AGN.

\subsection{Modelling Extinction Curves for AGN}

Mathis, Rumpl \& Nordsieck (1977, MRN) suggested compositions and
grain size distributions for Galactic interstellar dust. A
possible realization of their model contains two sorts of dust
with differing optical properties: carbonaceous or ``graphite''
grains, making up 37.5\% and ``astronomical silicate'' making up
62.5\%. This dust composition reproduces the interstellar media
extinction curve fairly well.

Weingartner \& Draine (2001) developed the MRN model to construct
size distributions for carbonaceous and siliceous grain
populations in different regions of the Milky Way and Magellanic
Clouds.  They present distributions that reproduce the observed
extinction through various lines of sight.  The $\lambda$2175
feature is caused predominantly by {\em small} carbonaceous
grains.  Inspection of their results shows that extinction curves
such as we find can be naturally explained by a relative lack of
both small carbonaceous and small siliceous grains.

We modelled our AGN extinction curves using a computer code which
calculates extinction cross sections from Mie theory. Such a code
is given by Bohren \& Hufman (1983).  It is based on the solution
of the Maxwell equations for radiation scattering by a spherical
grain with defined radius $a$ and dielectric constant $\epsilon$.
A graphite grain has differing optical properties for radiation
propagating parallel or perpendicular to the symmetry axis of the
crystal. The model takes care of this difference by using two
different kinds of graphite grains, $G_{r_{\|}}$ and
$G_{r_{\bot}}$. Considering three spatial dimensions there has to
be twice as much $G_{r_{\bot}}$ as $G_{r_{\|}}$. Further details
of our procedures can be found in Goosmann (2002). In the standard
MRN model the number density $n(a)$ of grains with radius $a$
follows a $n(a) \sim a^{\alpha_s}$ law, with $\alpha_s = -3.5$,
minimum grain size $a_{min} = 0.005 \mu m$ and maximum grain size
$a_{max} = 0.250 \mu m$. Based on this grain size distribution MRN
were able to compute the standard extinction curve for our Galaxy.

We recalculated the standard extinction curve and then varied the
dust composition, grain size limits and $\alpha_s$.  We found that
the slope of the UV extinction curve tends to be determined by
$a_{min}$ whereas the level of $E(\lambda-V)$ in the UV varies
more with $\alpha_s$. In Fig. 6 we show a fit to our reddening
curve.  In order to achieve good agreement we lowered the
abundance of graphite to 15\% of the graphite/silicate mixture.
In addition to this we raised $\alpha_s$ to $-2.05$ and lowered
$a_{max}$ to $0.200 \mu m$.

A possible physical basis for this grain-size distribution is the
following: when a standard Galactic dust cloud is irradiated by
hard quasar radiation small grains are preferentially destroyed,
and larger grains are partly depleted. Therefore the upper limit
of the grain radii should go down. The lower limit stays the same
because the destroyed small grains are replaced by depleted larger
ones. Hence there is always a fraction of small grains.  However,
since larger grains have a better resistance against the
destruction processes, their relative abundance rises and so does
$\alpha$.

It is hard to explain why the fraction of carbonaceous dust is
lower than in the MRN model since graphite actually has a higher
sublimation temperature than silicate.  The hard quasar radiation
should preferably destroy silicate.  This might be a limitation of
our model where the only way to lower the $\lambda2175$ feature is
by decreasing the carbon abundance.  In reality there might be
another carbonaceous substance, such as PAH molecules, not
included in our modeling, which is predominantly producing the
$\lambda2175$ feature.

We have made no attempt to model the reddening curve we found for
radio-quiet AGNs because we suspect that the differences between
this and the radio-loud reddening curve are caused by additional
reddening further out in the host galaxy while the ``nuclear''
dust is similar. The dust further out would have more ``Galactic''
properties than the nuclear dust.

It should be noted that our ``nuclear'' reddening curve is for
external line of sight reddening of a single point source.  This
is quite different from the case of starburst galaxies (Calzetti,
Kinney, \& Storchi-Bergmann 1994) where the dust is mixed in with
multiple sources, each with different extinctions, and where
geometrical effects in the complex environment affect the shape of
the reddening curve. Those effects can eliminate the strong
spectral curvature and exponential cutoff otherwise mandated by
the usual reddening curves. In our case there are no exponential
cutoffs despite the extinction being in the {\it foreground}.

\subsection{Mechanisms for Depleting Small Grains}

There are many mechanisms that can selectively destroy small
grains.  Gas in the outer regions of dense clouds in our galaxy is
observed to have a higher relative abundance of larger grains
(Mathis 1990), and this is believed to be due to coagulation of
smaller grains.  Gas near an AGN is close to the sublimation
radius, and small grains can be destroyed by single-photon
heating. Absorption of a single X-ray photon will eliminate grains
of radius $< 10 \AA$ and deplete grains several times larger than
this (Laor \& Draine 1993). As we have already mentioned in
section 6.2, Siebenmorgen et al. (2004) have shown how soft X-rays
destroy PAH molecules.  Small grains have a shorter lifetime due
to thermal sputtering (Draine \& Salpeter 1979).  Smaller grains
can also be efficiently destroyed by acquiring a high positive
charge from the photoemission of electrons in the intense hard
radiation field of AGNs (Draine \& Salpeter 1979; Chang, Schiano,
\& Wolfe 1987). The high positive charge of large grains will also
suppress thermal sputtering.  If the AGN is not radiating far
below the Eddington limit, grains will be efficiently accelerated
away from the black hole. The ratio of radiative acceleration to
gravitational acceleration is $\propto a^{-1}$, where $a$ is the
grain radius, so smaller grains will be expelled faster (see
Draine \& Salpeter 1979). The faster moving grains will also be
subject to enhanced sputtering (Laor \& Draine 1993).

\subsection{Continuum Variability}

In section 3 we have presented evidence for similarity of the
average continuum shapes of a wide variety of AGNs. We are not,
however, arguing for a precisely standard spectral energy
distribution. It is widely reported that as AGNs vary,
$\alpha_{UVO}$ changes, being steeper in low states (e.g., Perola
et al. 1982). However, much of the observed effect derives from a
constant component from the host galaxy. Also contributing is the
SBB atomic feature, which responds to continuum changes with
reduced amplitude and a time delay. Korista \& Goad (2001)
recently showed that atomic continua can cover a much wider
spectral region than just the SBB.

On the other hand, small delays have been reported between the UV
flux and the optical (Collier et al.  1998, 2001; Kriss et al.
2000; Collier 2001, Oknyanskij et al. 2002; Gaskell et al., 2004).
While there have been cases of reported steepening of the UV
spectrum as AGNs have varied, it is notable that in the case of
Fairall 9 {\em the UV continuum slope stayed the same} at $0.46
\pm 0.11$ while the UV flux changed by a factor of over 20
(Clavel, Wamsteker, \& Glass 1989).

Where the continuum does steepen as it fades, it is possible that
the extinction has changed (Barr 1982).  If this is the case it
would be due to compact dust clouds moving across the line of
sight.  There is support for this idea from the frequently-seen
variations in X-ray column densities on timescales of less than a
year (e.g., Risaliti, Elvis, Nicastro 2002), and the temporal
variations in the position angles of the polarizations of broad
lines in AGNs (e.g., Goodrich 1989, Martel 1998) which imply
motions of the scattering dust clouds near the broad line region.
More directly, Goodrich (1989) showed that the Balmer decrements
and continuum slopes change simultaneously in a way consistent
with varying reddening.

However, changing extinction is far from the whole story of
variability, since the X-rays vary, and often correlate with the
optical/UV (see Gaskell \& Klimek 2003 for review).

\subsection{Absorption Lines}

Dust is associated with gas, so if optically steep AGN spectra are
caused by reddening then we would expect that ``red'' AGNs would
show stronger absorption.  Wills et al. (2000) have indeed found
that AGNs with strong UV absorption lines have redder optical-UV
continua and either weaker or flatter soft X-ray continua.  Baker
et al. (2002) in their study of associated C\,IV absorption also
found that heavily-absorbed quasars are systematically redder and
they found that C IV associated absorption is found preferentially
in steep-spectrum and lobe-dominated quasars. The Wills et al.
(2000) and Baker et al. (2002) results provide additional
independent support the idea that changes in continuum shape are
due to dust.

\section{CONCLUSIONS}

If our reasoning is correct, we can draw several important
conclusions about AGNs.

1.  The reddening curve for radio-loud AGNs has a fairly universal
and unprecedented shape, being quite flat in the UV but selective
in the optical.

2.  The observed reddening of radio-quiet quasars is somewhat more
selective in the UV relative to the optical, compared with
radio-loud AGNs but this could well be because of additional dust
far out in the host galaxy.

3.  Even though most well-studied quasars show no UV curvature,
much of their radiated energy is in fact absorbed.

4. The flat nuclear reddening curve in the UV can be straightforwardly
explained by grain destruction depleting the number of small
grains, a notion consistent with many arguments in the literature.

5.  The normalization (average amount of extinction) for
lobe-dominant quasars on average decreases with radio core
dominance, as anticipated by Baker et al (1997) and Baker and
Hunstead (1995). There is powerful evidence from radio astronomy
that this means that absorption increases with line-of-sight
inclination to the radio jets, as expected qualitatively in the
standard ``unified model'' (Wills 1999).

6.  Compact Steep-Spectrum (CSS) radio quasars have the same type
of extinction curve and high reddening.

7. The continuum shapes of all but the bluest quasars are affected
by reddening, and are intrinsically similar to those of the bluest quasars.

8. The average reddening decreases substantially with increasing luminosity.

9. The luminosity dependence of extinction explains most or all of
the luminosity dependence of $L_X/L_{opt}$, and the variation of
this dependence with the choice of optical wavelength.

10. Differences in reddening are also a major factor in the
difference in $L_X/L_{opt}$ between radio-loud and radio-quiet
AGNs.

11.  If they are correct, many of these results are of vital
importance for demographic studies (intrinsic luminosity density
of AGNs in the universe), and for modelling the crucial Big Blue
Bump continuum component of AGN.

The most satisfying thing about the results from this paper are
they are readily and robustly testable.  The nearly unequivocal
prediction is that radio-loud quasars must be generally powerful
thermal radiators. This is especially true for those with steep
optical slope, despite their general lack of exponential UV
cutoffs as expected in the case where small grains are present.
While the exact infrared luminosities are model-dependent, only
predictable statistically, and far beyond the scope of this paper,
they should generally be energetically very important, and often
dominant.  The one obvious caveat is that the optical/UV radiation
may be intrinsically emitted anisotropically, as has been argued
on other grounds by Miller, Goodrich, \& Mathews (1991) among
others.  This could reduce the amount of re-processed radiation.

\acknowledgments

We are grateful to Mark Bottorff, Patrick Hall, Ari Laor, John
Mathis, Dick McCray, Rita Sambruna, Joe Shields, Joe Weingartner,
\& Adolf Witt for useful comments and discussion.  This research
was supported in part by NSF grant AST-0098719 and by the
Hans-B\"{o}ckler-Stiftung.

\appendix

\section{AN AGN EXTINCTION CURVE}

In Table 3 we give an average extinction curve for the optical and
the UV waveband. It is based on Fig. 1 and we have tabulated it at
the wavelengths of major emission lines as well as the B and the V
band. Note that values for wavelengths corresponding to $x$
outside the range 1.6--8 $\mu m^{-1}$ are extrapolated.

The reddening curve can be represented analytically by the
following functions of $x = \lambda^{-1}$ in $\mu m^{-1}$:

$
\begin{array}{ll}
 A_\lambda/A_V(x) = -0.8175+1.5848x-0.3774x^2+0.0296x^3, &
 1.6 \mu m^{-1} \leq x < 3.69 \mu m^{-1}, \\
 A_\lambda/A_V(x) = 1.3468+0.0087x, &
 3.69 \mu m^{-1} \leq  x \leq 8 \mu m^{-1}.
\end{array}
$

The analytic fit is shown in Fig. 6 after conversion to
$\frac{E(\lambda-V)}{E(B-V)}$ using $R_V = 4.15$. This $R_V$ value
is an average of the optical/UV and the far UV values derived
following CCM89 from our radio-loud reddening curves (see section
2.4).

\clearpage

\begin{deluxetable}{lccc}
 \tablewidth{0pt}
 \tablecaption{Estimates of the Ratio of Total to Selective Extinction}
 \tablehead{
 \colhead{Data sets} & \colhead{IR-extrap.} & \colhead{opt./near-UV} & \colhead{far-UV}}
 \startdata
   $\Re < 0.1$ to $\Re \geqslant 1$ & $5.30\pm0.15$ & $5.55\pm0.15$ & $5.40\pm0.20$ \\
   CSS  to $0.1 \leqslant \Re < 1 $ & $4.95\pm0.15$ & $4.95\pm0.15$ &  $4.80\pm0.15$ \\
   LBQS to $\Re \geqslant 1$ & $3.70\pm0.25$ & $3.80\pm0.20$ & $3.20\pm0.20$
 \enddata
\end{deluxetable}

\clearpage

\begin{deluxetable}{ccc}
 \tablewidth{0pt}
 \tablecaption{Effect of Host Galaxy Contamination on Inferred Reddening}
 \tablehead{
 \colhead{AGN/Host} & \colhead{E(B-V)$_{opt}$} & \colhead{E(B-V)$_{UV-opt}$}}
 \startdata
    10 & 0.02 & 0.04\\
    5 & 0.04 & 0.07\\
    1 & 0.12 & 0.31\\
    0.5 & 0.17 & 0.54\\
    0.1 & 0.25 & 1.50
 \enddata
\end{deluxetable}

\clearpage

 \begin{deluxetable}{lcc}
 \tablewidth{0pt}
 \tablecaption{AGN Extinction at Major Emission Lines and Wavebands}
 \tablehead{
 \colhead{Line} & \colhead{$\lambda$ [\AA]} & \colhead{$A_\lambda/A_V$}}
 \startdata

    O I & 8446 & 0.579\\
    H $\alpha$ & 6563 & 0.826\\
    He I & 5876 & 0.932\\
    V band & 5500 & 0.994\\
    H $\beta$ & 4861 & 1.103\\
    He II & 4686 & 1.133\\
    B band & 4400 & 1.182\\
    He I & 4471 & 1.170\\
    H $\gamma$ & 4340 & 1.193\\
    Mg II & 2798 & 1.377\\
    C II] & 2326 & 1.384\\
    C III] & 1909 & 1.392\\
    He II & 1640 & 1.400\\
    C IV & 1549 & 1.403\\
    O IV]/Si IV & 1400 & 1.409\\
    O I & 1304 & 1.414\\
    N V & 1240 & 1.417\\
    Ly $\alpha$ & 1216 & 1.418\\
    O IV & 1034 & 1.431
 \enddata
\end{deluxetable}

\clearpage

\begin{figure} \plotone{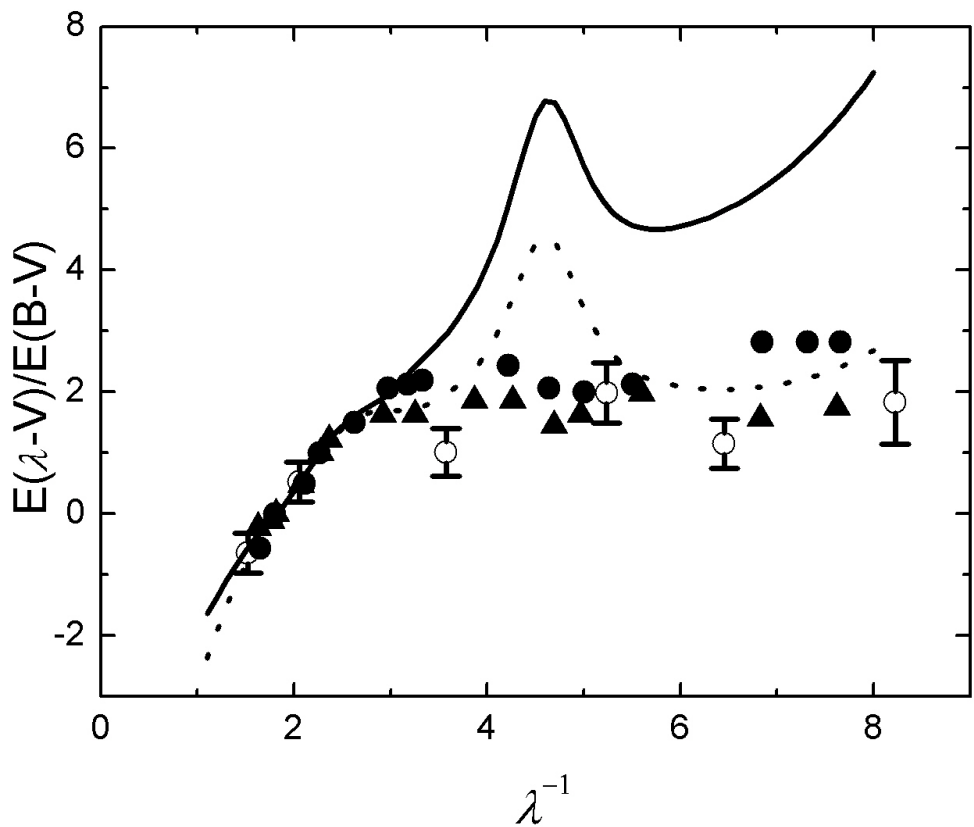} \caption{Reddening curves (in magnitudes vs.
$\mu$m$^{-1}$) based on the Baker \& Hunstead (1995, 1996) data
subsets for composite spectra.  Filled triangles are from
comparing $\Re \geqslant 1$ with $\Re \leqslant 0.1$ and filled
circles are from comparing $0.1 \leqslant \Re < 1$ with CSS. The
open circles represent average BLR extinction values between
face-on $(\Re \geqslant 1)$ and edge-on ($\Re < 0.1$, CSS)
objects. Theoretical reddening curves derived from CCM89 for $R_V
= 5.3$ and $R_V = 3.1$ are shown as dashed and solid curves
respectively. \label{fig1}} \end{figure}


\begin{figure} \plotone{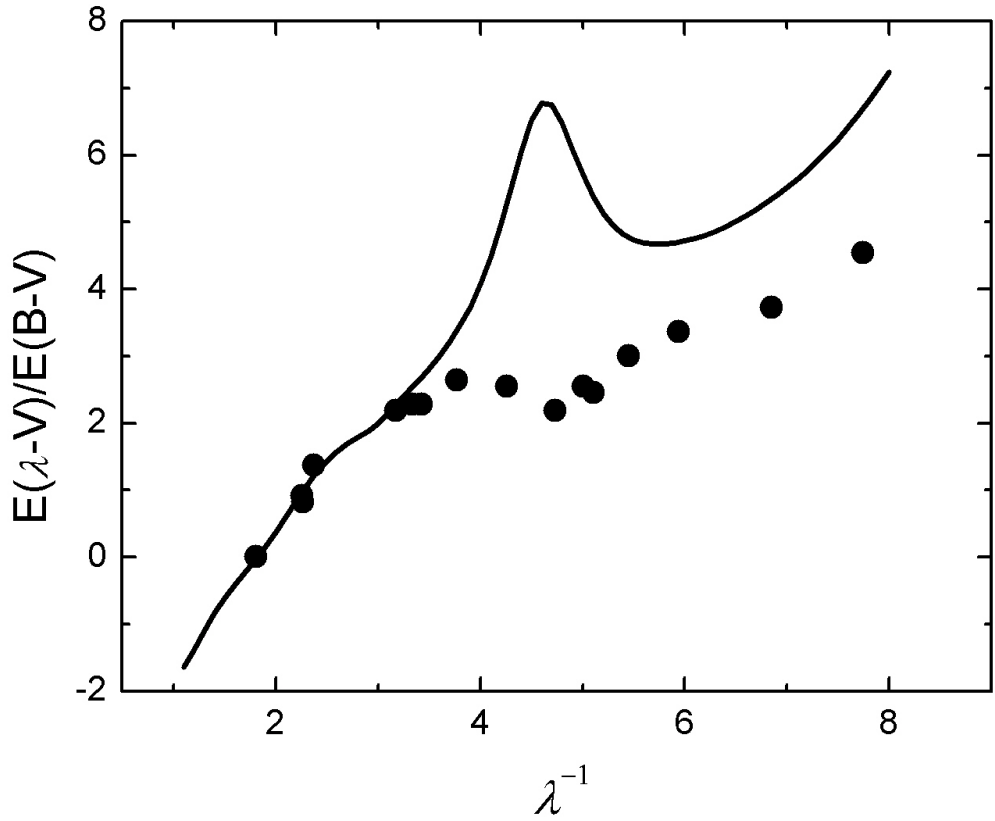} \caption{Reddening
curve (units as in Fig. 1) based on the Baker \& Hunstead data for
the composite spectrum $\Re \geqslant 1$ compared to the LBQS
sample (circles). Computed reddening curve derived from CCM89 for
$R_V = 3.1$ (solid curve). \label{fig2}} \end{figure}


\begin{figure} \plotone{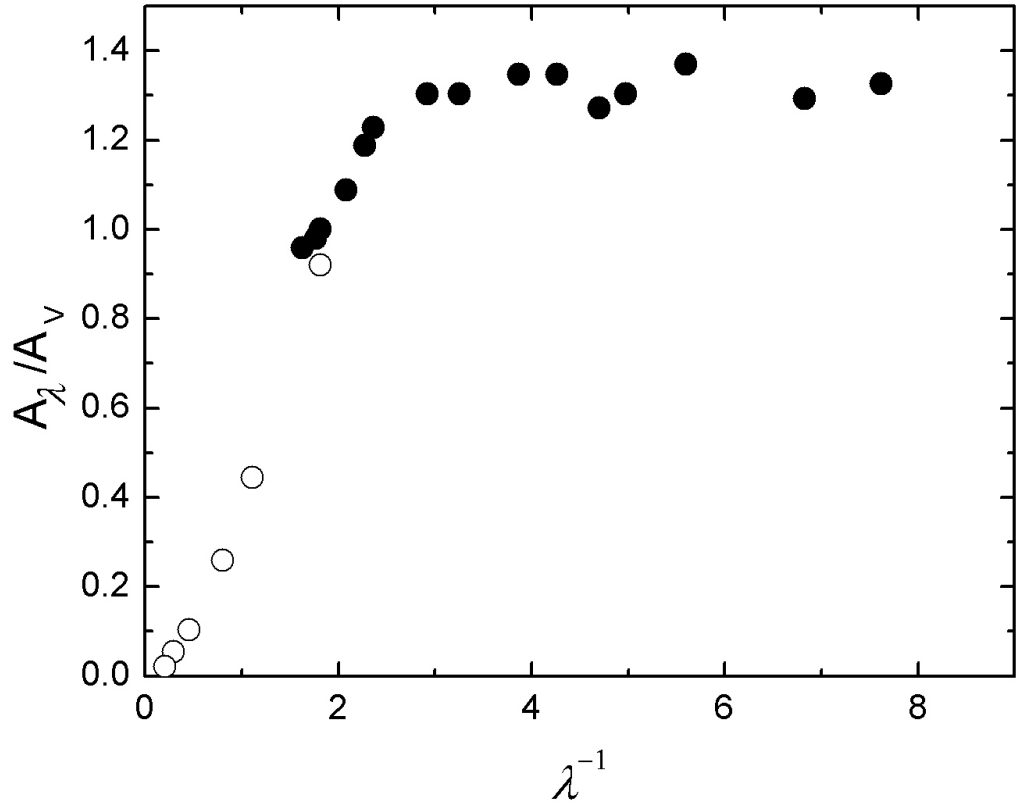} \caption{Optical and UV reddening
curve, in magnitudes as a function vs. inverse wavelength
($\mu$m$^{-1}$), between the composite spectra $0.1 \leqslant \Re
< 1$ and CSS (filled circles) with an extrapolation into the IR
based on Rieke \& Lebofsky (open circles). \label{fig3}}
\end{figure}


\begin{figure} \plotone{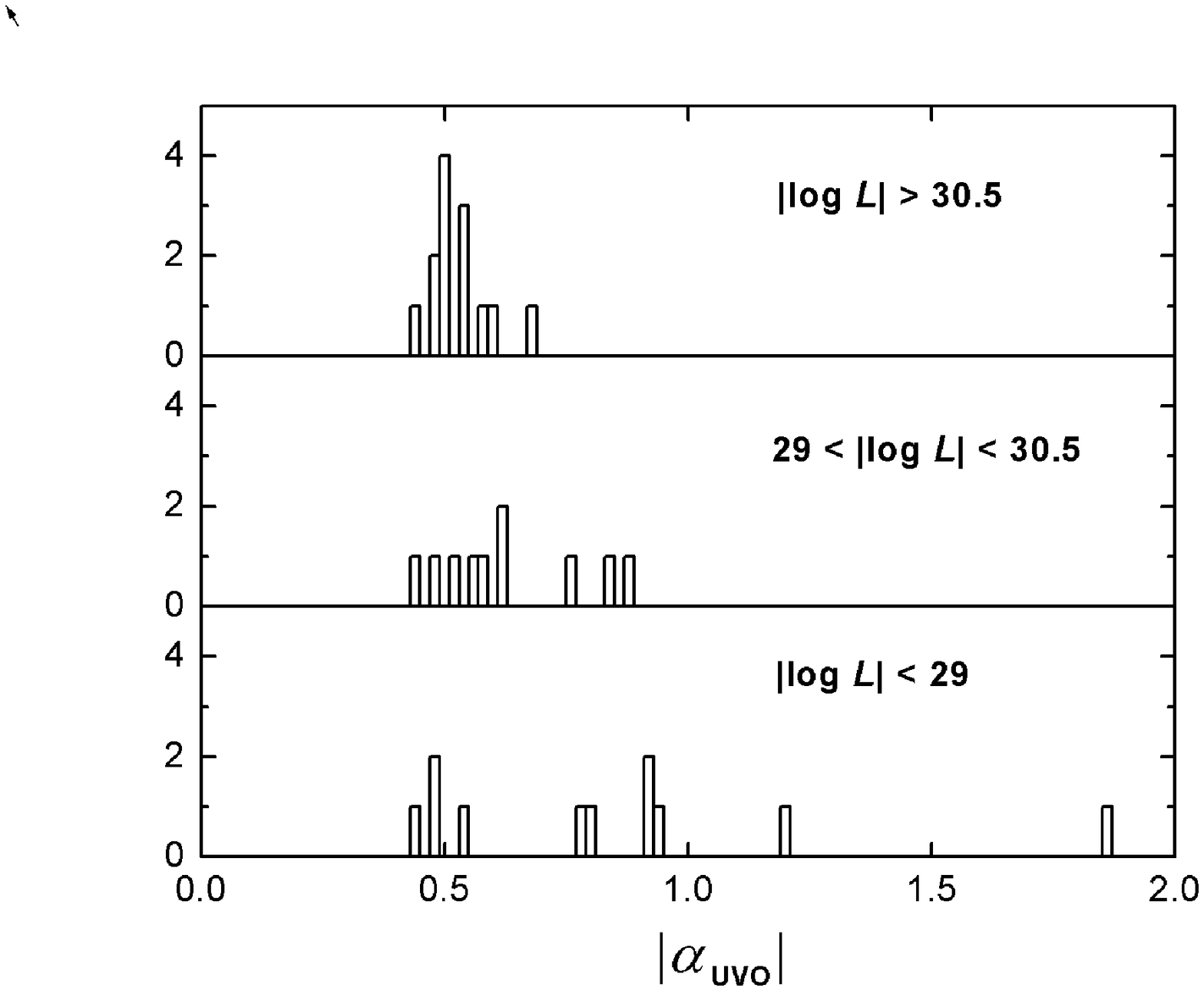} \caption{Number distribution
of $\alpha_{UVO}$ for AGNs in three different monochromatic
optical luminosity ($L_{4200}$) ranges, $\vert logL \vert > 30.5$
(a), $29 < \vert logL \vert < 30.5$ (b) and $\vert logL \vert <
29$ (c) \label{fig4}}
\end{figure}


\begin{figure} \plotone{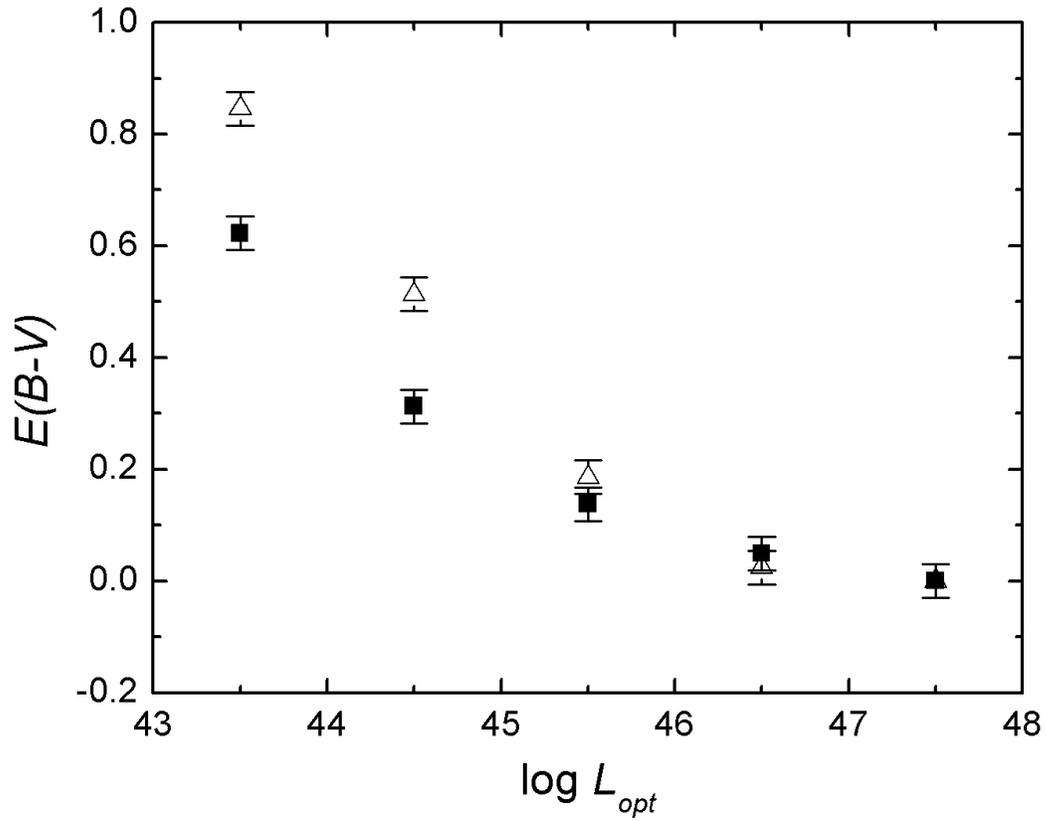} \caption{The mean
reddenings, (E(B-V)), inferred using the extinction curve in Fig.
1, plotted as a function of observed optical luminosity of the
AGNs (in ergs s$^{-1}$). The solid squares are reddenings derived
from the $\lambda$1600 to $\lambda$4200 spectral index; the open
triangles are the reddenings derived from the $\lambda$4200 to
$\lambda$7000 spectral index. The trends are similar using the
optical and UV diagnostics.}

\end{figure}


\begin{figure} \plotone{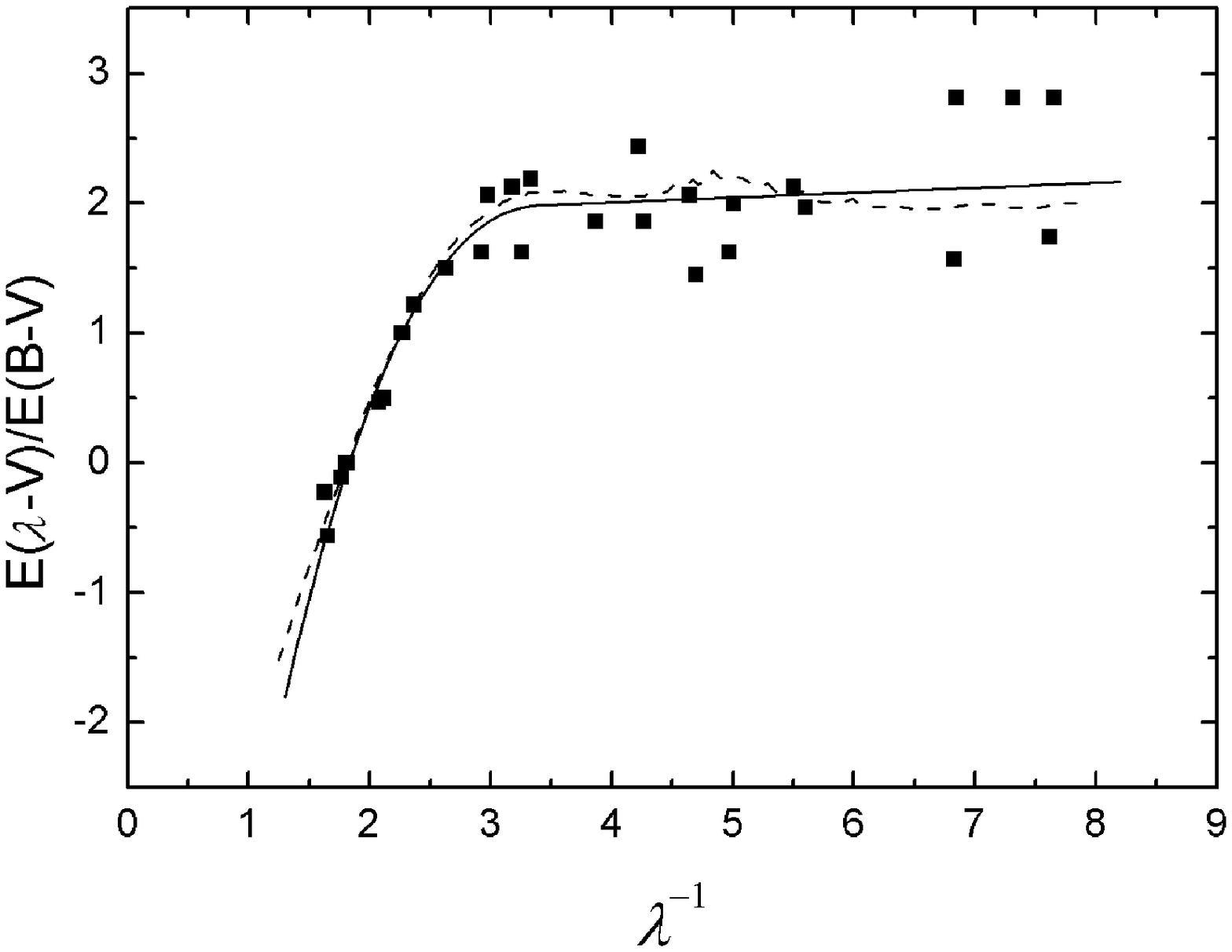} \caption{AGN extinction curves (squares)
plotted together with a model curve derived from Mie theory
computations (dotted line, see section 6.2) and the analytical fit
given in the Appendix (solid line) \label{fig6}}
\end{figure}



\begin{thebibliography}{}
 \bibitem{} Antonucci, R. R. J. 1993, \araa, 31, 473
 \bibitem{} Antonucci R. R. J. 2002, in Astrophysical
    Spectropolarimetry, eds. J. Trujillo-Bueno, F.
    Moreno-Insertis, \& F. Sanchez (Cambridge: Cambridge Univ. Press), p. 151.
 \bibitem{} Antonucci R. R. J.
    et al. 1996, \apj, 472, 502
 \bibitem{} Avni, Y. \& Tananbaum, H. 1982, \apj, 262, L17
 \bibitem{} Baker, J. C., et al. 1997,  MNRAS 286, 23
 \bibitem{} Baker, J. C., \& Hunstead, R. W. 1995, \apj, 452, L95
 \bibitem{} ------------ 1996a, \apj, 461, L59 (Erratum)
 \bibitem{} ------------ 1996b, \apj, 468, L131 (Erratum)
 \bibitem{} Baker, J. C., Hunstead, R.W., Athreya, R. M., Barthel, P. D.,
    de Silva, E., Lehnert, M. D., \& Saunders, R. D. E. 2002, \apj, 568, 592
 \bibitem{} Barr, P. 1982, in ESA 3rd European IUE Conf., p. 559
 \bibitem{} Bechtold, J., Shields, J., Rieke, M., Ji, P., Scott J.,
    Kuhn, O., Elvis, M., \& Elston, R. 1997, in ASP Conf. Ser. 113,
    Emission Lines in Active Galaxies: New Methods and Techniques, ed.
    B. M. Peterson, F.-Z. Cheng, \& A. S. Wilson (San Francisco: ASP), 123
 \bibitem{} Bohren C. F. \& Hufman D. R., ``Absorption and
    Scattering of light by small particles'' (New York: Wiley \& Sons)
 \bibitem{} Brinkman, W., Yuan, W, \& Siebert, J 1997, A\&A, 319, 413
 \bibitem{} Burstein, D., \& Heiles, C. 1982 \aj, 87, 1165
 \bibitem{} Calzetti, D., Kinney, A. L., \& Storchi-Bergmann, T.
    1994, \apj, 429, 582.
 \bibitem{} Cardelli, J. A., Clayton G. C., \& Mathis, J. S. 1989,
    \apj, 345, 245 (CCM89)
 \bibitem{} Cheng, F. H., Gaskell, C. M., \& Koratkar, A. P. 1991,
    \apj, 370, 487
 \bibitem{} Clavel, J., Schulz, B., Altieri, B., Barr, P., Claes,
     P., Heras, A., Leech, K., Metcalfe, L., \& Salama, A. 2000, A\&A,
     357, 839
 \bibitem{} Clavel, J., Wamsteker, W., Glass, I. S., 1989, \apj, 337, 236
 \bibitem{} Chang, C. A., Schiano, A. V. R., \& Wolfe, A. M. 1987,
    ApJ, 322, 180
 \bibitem{} Collier, S. J. 2001, \mnras, 325, 1527
 \bibitem{} Collier, S. J., et al. 1998, \apj, 500, 162
 \bibitem{} Collier, S. J., et al. 2001, \apj, 561, 146
 \bibitem{} Crenshaw, D. M., Kraemer, S. B., Bruhweiler, F. C., \& Ruiz, J.
     R. 2001, ApJ, 555, 663
 \bibitem{} Crenshaw, D. M., Kraemer, S. B., Turner, T. J., Collier, S.,
    Peterson, B. M., Brandt, W. N., Clavel, J., George, I. M., Horne, K.,
    Kriss, G. A., Mathur, S., Netzer, H., Pogge, R. W., Pounds, K. A.,
    Romano, P., Shemmer, O., \& Wamsteker, W. 2002,  ApJ, 566, 187
 \bibitem{} De Zotti, G., \& Gaskell, C. M. 1985, A\&A, 147, 1
 \bibitem{} Draine, B. T. \& Salpeter, E. E. 1979, ApJ, 231, 77
 \bibitem{} Dultzin-Hacyan, D., and Ruano, C. 1996, A\&A305, 719
 \bibitem{} Fischera, Jg., Tuffs, R. J., V\"{o}lk, H. J. 2002, A\&A, 395, 189
 \bibitem{} Fizpatrick, E. L., \& Massa, D. 1988, \apj, 328, 734
 \bibitem{} Francis, P. J., Hewett, P. C., Foltz, C. B., Chaffee, F. H. 1992,
    \apj, 398, 476
 \bibitem{} Gaskell, C. M., Doroshenko, V. T., Klimek, E. S., Crowley, K. A.,
     George, T. A., Grove, R., Hedrick, C. H., Hiller, M. E., Peterson, B. W., \& Poulsen,
     M. A. et al., 2004, in preparation
 \bibitem{} Gaskell, C. M. \& Klimek, E. S. 2003, A\&ApT, 23, 22, 661.
 \bibitem{} Goodrich, R. W. 1989, \apj, 340, 190
 \bibitem{} Goosmann, R. W. 2002, Diplomarbeit thesis, Univ.
     Hamburg
 \bibitem{} Gordon, K. D., Hanson, M. M., Clayton, G. C., Rieke, G. H., \&
    Misselt, K. A. 1999, \apj, 519, 165
 \bibitem{} Gordon, K. D., \& Clayton, G. C. 1998, \apj, 500, 816
 \bibitem{} Grandi, S. A. 1983, \apj, 268, 591
 \bibitem{} Grossan, B., Remillard, R. A., Bradt, H. V., Brissenden, R. J., Ohashi, T., \& Sakao,
 T. 1996, ApJ, 457, 199
 \bibitem{} Hall, P. S. et al., 2002, \apjs, 141, 267
 \bibitem{} Imanishi, M. 2001, AJ, 121, 1927
 \bibitem{} Kembhavi, A. 1997, \mnras, 264, 683
 \bibitem{} Korista, K. T. \& Goad, M. R. 2001, \apj, 553, 695
 \bibitem{} Kriss, G. A., Peterson, B. M., Crenshaw, D. M.,
    Zheng, W.  2000, 535, 58

 \bibitem{} Kriss, G. A. \& Canizares, C. R. 1985, ApJ, 297, 177
 \bibitem{} Krolik, J. H. \& Begelman, M. C. 1988, \apj, 329, 702
 \bibitem{} Laor, A., Draine, B. T, 1993, ApJ, 402, 441
 \bibitem{} Maccacaro, T., Perola, G. C., \& Elvis, M. 1982, ApJ, 257, 47
 \bibitem{} Maiolino, R., Marconi, A., \& Oliva, E. 2001a, A\&A, 365, 37
 \bibitem{} Maiolino, R., Marconi, A., Salvati, M., Risaliti, G., Severgnini,
    P., Oliva, E., La Franca, F., \& Vanzi, L. 2001b, A\&A, 365, 28
 \bibitem{} Malkan, M. A. 1984, in ``X-Ray and UV Emission from Active
     Galactic Nuclei'', Max-Planck Inst. Reports, Vol. 184, p. 121
 \bibitem{} Maoz, D., Korista, K. T., Shapovalova, A. I., Shields, J. C.,
    Smith, P. S., Thiele, U., Wagner, R. M. 1993, \apj, 404, 576
 \bibitem{} Martel, A. R. 1998, \apj, 508, 657
 \bibitem{} Mathis, J. S. 1990, \araa, 28, 37
 \bibitem{} Mathis, J. S., Rumpl, W., \& Nordsiek, K. H. 1977, \apj, 217, 425 (MRN)
 \bibitem{} McKee C. F., \& Petrosian V. 1974, \apj, 189, 17
 \bibitem{} Miller, J. S., Goodrich, R., \& Mathews, W. G. 1991, \apj, 378, 47.
 \bibitem{} Misselt, K. A., Clayton, G. C., \& Gordon, K. D. 1999, \apj, 515, 128
 \bibitem{} Mushotzky, R. F. \& Wandel, A. 1989, \apj, 339, 674
 \bibitem{} Netzer H. 1985, \apj, 289, 451
 \bibitem{} Netzer H., \& Davidson K. 1979, \mnras, 187, 871
 \bibitem{} Netzer H., Brotherton M. S., Wills B. J., Han M. S., Wills D.,
    Baldwin J. A., Ferland G. J., \& Browne, I. W. A. 1995, \apj, 448, 27
 \bibitem{} Ogle, P. M., Brookings, T., Canizares, C. R., Lee, J. C., \& Marshall, H.
    L. 2003, A\&A, 402, 849.
 \bibitem{} Oknyanskij, V. L., Horne, K., Lyuty, V. M., Sadakane, K., Honda, S., \& Tanabe,
    S.in Active Galactic Nuclei, from Central Engine to Host Galaxy, eds. S. Collin, F.
    Combes and I. Shlosman. ASP Conf. Ser, Vol. 290, p. 119
 \bibitem{} Perola, C. et al. 1982, \mnras, 200, 293
 \bibitem{} Pitman K. M., Clayton G. C., \& Gordon K. D. 2000, \pasp, 112, 537
 \bibitem{} Reichert, G. A., Mason, K. O., Bowyer, S. \& Thorstensen, J. R. 1982, ApJ, 260, 437
 \bibitem{} Reichert, G. A., Mushotzky, R. F., Holt, S. S., \& Petre, R.
    1985, ApJ, 296, 69
 \bibitem{} Rieke G. H., \& Lebofsky M. J. 1985, \apj, 288, 618
 \bibitem{} Richards, G. T. et al., 2003, AJ, 126, 1131
 \bibitem{} Risaliti, G., Elvis, M., \& Nicastro, F. 2002, \apj, 571,
    234
 \bibitem{} Sambruna, R. M., Eracleous, M., \& Mushotzky, R. F. 199, ApJ, 526,
    60
 \bibitem{} Schmitt, H. R., Kinney, A. L., Calzetti, D., Storchi-Bergmann, T.
    1997, AJ, 114, 592
 \bibitem{} Shastri, P. 1991, \mnras, 249, 640

 \bibitem{} Siebenmorgen, R., Krügel, E., \& Spoon, H. W. W. 2004,
 A\&A, 414, 123

 \bibitem{} Snedden, S. A. \& Gaskell, C. M. 2004, ApJ, submitted.

 \bibitem{} Wandel, A. \& Petrosian, V. 1988, \apj, 329, L11
 \bibitem{} Ward M. J., Elvis M., Fabbiano G., Carleton N. P., Willner S. P.,
    \& Lawrence A. 1987, \apj, 315, 74
 \bibitem{} Ward, M. J. \& Morris, S. L. 1984, MNRAS, 207, 867
 \bibitem{} Weingartner J. C., \& Draine B. T. 2001, \apj, 548, 296
 \bibitem{} Wills, B. J. 1999, in Quasars and
    Cosmology, ASP Conference Series 162, ed. G. J. Ferland \& J. A.
    Baldwin. (San Francisco: Astron. Soc. Pacific), p. 101
 \bibitem{} Wills, B. J., Shang, Z.-H, \& Yuan, J. M. 2000, New
    Ast. Rev. 44, 511
 \bibitem{} Zamorani, G., Henry, J. P., Maccacaro, T., Tananbaum, H., Soltan,
    A., Avni, Y., Liebert, J., Stocke, J., Strittmatter, P. A., Weymann, R. J.,
    Smith, M. G., \& Condon, J. J. 1982, ApJ, 245, 357
\end{thebibliography}
\end{document}